\documentclass[preprint,12pt]{aastex}

\def\msun{{\rm\,M_\odot}} 
\def\lsun{{\rm\,L_\odot}}

\newcommand{\etal}{et al.\ }

\newcommand{\lya}{Ly$\alpha$ }

\def\h2{${\rm\,H_2}$}

\slugcomment{Submitted to ApJ}

\begin{document}

\title{Very Large H II Regions Around Proto-Clusters of Galaxies During Reionization}

\author{Renyue Cen\altaffilmark{1}}

\altaffiltext{1} {Princeton University Observatory, 
Princeton University, Princeton, NJ 08544; cen@astro.princeton.edu}

\accepted{ }

\begin{abstract}

The physical basis for the belief that
the abundant old dwarf galaxies seen in present-day galaxy 
clusters formed before reionization of the universe is very compelling,
because (1) the observed faint end slope of the galaxy luminosity function 
ascends from a shallower slope at brighter magnitudes
and tracks that of dark matter halo mass function,
and (2) that steep power-law slope is observed
to extend all the way to galaxies of inferred velocity dispersion of $\sim 10$km/s.
We then show that the number of ionizing 
photons emitted by these dwarf galaxies 
should be able to produce an H II region at least as large as
each proto-cluster region, occupying at least $20\%$ of the entire IGM volume
at $z\ge 6$.
A more likely scenario is that our estimate
of the ionizing photon production rate based on these dwarfs 
under-estimates the true rate from proto-clusters 
by a factor of $\sim 5$ for a variety of physically plausible reasons,
in which case a fully self-consistent picture emerges.
Specific implications include
(1) Strong clustering of sources would collectively 
produce very large individual H II regions of sizes
up to $\sim 100$Mpc before overlapping with others,
having important implications for upcoming radio and CMB observations.
There should be large spatial fluctuations of 
neutral fraction in the intergalactic medium up to 
cluster scales of $\sim 100$Mpc.
We give an H II region size function.
(2) The ionizing sources at $z\sim 6$ are expected to be
highly biased with a bias factor $\ge 15$.
(3) Large H II regions would enable largely unattenuated transmission of
flux of a significant number of sources prior to reionization.
When more sensitive observations become available,
one should expect to see clusters of fainter sources
in the vicinity of known galaxies at $z>6$.

\end{abstract}


\section{Introduction}

The entire cosmological reionization process may be 
extended in time and quite complex
(Kogut \etal 2003), likely with a few twists and turns
(e.g., 
Cen 2003; Wyithe \& Loeb 2003; Haiman \& Holder 2003; Ricotti \& Ostriker 2004).
We just began to understand this process.
The evidence for the completion of the final reionization episode
at $z\sim 6$ is, however, firmly established
by the recent observations of high redshift 
quasars from the Sloan Digital Sky Survey (SDSS)
(e.g., Fan \etal 2001; Becker \etal 2001; Barkana 2002; 
Cen \& McDonald 2002; White \etal 2003).

What is still unclear is what sources are responsible 
for ionizing the universe at $z\sim 6$ and
a variety of scenarios may be imagined,
although no fully satisfactory scenario has been put forth.
In this {\it Letter} 
we suggest that a large portion, if not all, 
of the required ionizing sources 
may be the old, faint dwarf galaxies,
already observed in local clusters of galaxies
from the RASS-SDSS cluster survey (Popesso \etal 2005).
We present physical reasoning why these local sources
were formed prior to reionization and
show the star formation rate from them 
is capable of ionizing at least a significant 
fraction of the universe at $z\sim 6$ and
most likely the entire intergalactic medium (IGM).

\section{Dwarf Galaxies in Clusters of Galaxies and Reionization}

The RASS-SDSS galaxy cluster survey (Popesso \etal 2005)
convincingly demonstrates that the faint end 
($M_z=-13$ to $-18$) of the galaxy luminosity
function (LF) for early-type galaxies 
has a slope $\alpha$ about and somewhat
steeper than $-2.0$ in all three SDSS bands 
($\alpha=-2.01\pm 0.11$ in $r$,
$\alpha=-2.03\pm 0.08$ in $i$,
and $\alpha=-2.05\pm 0.09$ in $z$ band)
(Figure 9 of Popesso \etal 2005).
For simplicity we will just use $z$ band for further analyses.

Let us first check
what kind of galaxies $M_z=-13$ corresponds to.
For that we will draw upon our present knowledge of dwarf galaxies
in the Local Group (LG), where they are best studied.
Two well studied relevant dwarfs in the LG 
are Fornax and Sagittarius, which have $M_V=-13.2$ and $-13.4$,
respectively (Table 4 of Mateo 1998).
Using the magnitude conversions in Table 3 of Frei \& Gunn (1994)
for type E galaxies, we obtain their respective 
$z$ band absolute magnitude $M_z=-14.26$ and $-14.46$.
The observed 1-d velocity dispersions 
for Fornax and Sagittarius
are $10.5\pm 1.5$km/s and 
$11.4\pm 0.7$km/s, respectively (Figure 7 of Mateo 1998). 
While one cannot directly translate LG dwarfs to cluster dwarfs,
it seems reasonable to assume that the faint limit at $M_z\sim -13$ observed
by Popesso \etal (2005) may correspond to galaxies 
of velocity dispersion close to $10$km/s. 
Kravtsov, Gnedin \& Klypin (2004) presented a quite interesting 
case for the observed LG dwarfs to have significantly
more massive progenitors in the past.
They suggest that the velocity dispersion of the progenitors 
at the epoch of major star formation may be a
factor of $2$ larger.
We thus propose that the observed dwarfs with $M_z\sim -13$ 
have velocity dispersions at formation of $10-20$km/s.

Having plausibly established the physical nature of these galaxies,
we can now ask the following question.
Did these systems form prior or subsequent to reionization?
The effect of photo-heating due to reionization
on the suppression of dwarf galaxy
formation is well demonstrated (Bullock, Kravtsov, \& Weinberg 2000;
Somerville 2002; Benson \etal 2002),
indicating that galaxy formation in halos with velocity dispersions
less than $\sim 30$km/s is greatly suppressed and
the effect extends well into larger halos
(Thoul \& Weinberg 1996; Quinn, Katz, \& Efstathiou 1996; Gnedin 2000).
Thus, it would be extremely unlikely 
(if not impossible) for the faint galaxies with $\sigma=10-20$km/s
to maintain a power-law slope of $\sim -2$, 
which is the slope of the underlying dark matter halo mass function
(Press \& Schechter 1974), if they were formed subsequent to reionization.
This statement would remain valid even if we have under-estimated
the velocity dispersion of these systems by a factor of a few.
This is primarily because it would be extremely unlikely 
to produce a power-law luminosity function of slope $-2$
for halos of estimated mass in the range of $10^8-10^{10}\msun$,
while the slope of the halo mass function at faint end is also $-2$,
when the Jeans mass of post-reionization gas 
is raised to $2\times 10^{10}$ for $T=10^4~$K.
If, for some reason, gas was stored in these systems prior to reionization
then later formed stars subsequent to reionization,
as Shaviv \& Dekel (2003) showed, photo-heating
would boil gas out of potential wells of halos 
with $\sigma\sim 30$km/s.
Moreover, the faint end slope of $\alpha \sim -2$ from $M_z=-18$ to $M_z=-13$
ascends from a much shallower ($\alpha\sim -0.76$) slope
at brighter magnitudes.
Since photo-heating induced suppression of star formation
is progressively greater for smaller galaxies,
the upturn of the luminosity function at low end
closely tracking the underlying mass function
is just the opposite of what one would expect if they
form subsequent to reionization.
We conclude that the faint galaxies displaying a power-law LF slope
of $\sim -2$ must have formed before photo-heating due to 
reionization has significantly affected them.

Additional credible evidence that supports this picture 
comes from the observed abundance differences between 
cluster dwarfs and dwarfs in lower density regions
(Trentham \& Tully 2002).
Tully \etal (2002) were the first to suggest
that reionization may be able to explain 
this abundance difference.
They show that dwarf halos in clusters
are able to collapse before reionization to form stars,
whereas dwarf halos in lower density regions collapse after
the reionization and were unable to accrete 
a significant amount of gas to form visible galaxies,
hence the lack of them.
Our work extends this idea and 
relates these cluster dwarfs directly
to sources that reionize the universe at $z\sim 6$,
in a self-consistent way.

The next question then arises:
If these faint galaxies in present clusters were formed
at epochs prior to $z\sim 6$,
can they be responsible for producing 
the majority of ionizing photons for the final reionization at $z\sim 6$?
We show below that they can indeed produce a large 
number of ionizing photons to at least significantly ionize
the universe at $z\sim 6$.

Using $M_z=4.51 -2.5\log(L/\lsun)$ (Kauffmann \etal 2003),
we find the luminosity $L=1.01\times 10^7\lsun$ for $M_z=-13.0$.
Using stellar mass to luminosity ratio in z-band $(M/L)_z=0.5-0.6$ for 
$M_z=-14.0$ to $-18.0$ (Figure 14 of Kauffmann \etal 2003)
we find the stellar mass 
to be approximately $M_s=5.56\times 10^6\msun$ for $M_z=-13.0$.
It is interesting to note, if star formation efficiency
is $10\%$, one would obtain total gas mass for each galaxy with $M_z=-13.0$
of $M_g=5.56\times 10^7\msun$ and total mass of 
of $M_t=3.5\times 10^8\msun$ corresponding to a velocity dispersion of $17$km/s at $z=6$.
This indicates that our assumption for the velocity dispersion
is self-consistent.
Integrating the LF function from 
$M_z=-18.0$ to $-13.0$ (lower right panel of Figure 9 in Popesso \etal 2005)
we obtain the total stellar mass contained
in these faint galaxies in the 69 sampled clusters of 
Popesso \etal (2005) to be $1.00\times 10^{13}\msun$.
For every baryon formed into stars with the standard Salpeter IMF
about $3500$ hydrogen ionizing photons are produced.
Assuming ionizing photon escape fraction of $f_{esc}$,
the total number of ionizing photons pumped into 
the IGM from all the faint galaxies
in these 69 clusters are equal to the number
of hydrogen atoms contained 
in $4.60\times 10^{15} (f_{esc}/0.1)\msun$ of gas.

Next, let us compute the total gas mass contained
in the 69 clusters in the RASS-SDSS sample.
Based on data presented in Popesso \etal (2004)
we estimate that the total mass of the 69 clusters 
is $2.5\times 10^{16}\msun$, yielding
the total gas mass contained in these
clusters of $4.0\times 10^{15}\msun$.
We see that the number of photons produced by the faint galaxies
[$4.60\times 10^{15} (f_{esc}/0.1)\msun$]
is about equal to the number of hydrogen atoms
in all these clusters, if $10\%$ of ionizing photons were able to 
escape from star-forming galaxies.
It thus seems likely that 
the ionizing photons produced by the known 
faint cluster galaxies alone
can at least ionize all the hydrogen in the proto-cluster regions.
Since approximately $20\%$ of all matter is in clusters of galaxies
today, hence, at face value, 
it appears that the known dwarf galaxies in clusters 
fall short by a factor of $\sim 5$ to produce 
one ionizing photon per hydrogen atom for the entire IGM.

Could we have underestimated the number of ionizing photons
from galaxies in proto-clusters?
There are at least six possible sources that may have caused
such an underestimation.
First, it should be noted that the majority 
of dark matter halos that
originally host these dwarfs galaxies at $z\ge 6$
most likely have disappeared by $z=0$, due to 
merging and disruption over the time.
The stellar components, presumably having collapsed by a larger factor,
are often thought to be more robust for survival
(e.g., Gao \etal 2004).
Nevertheless, it is unknown presently about the
survival rate of these galaxies 
and it is possible that a large fraction of such stellar systems
may have lost their separate entities in the course of time
since $z\sim 6$.
Second, the limits of $M_z=-18.0$ to $-13.0$ of the LF that
we use may have excluded some galaxies at brighter magnitudes.
How abundant the galaxies at $M_z<-18.0$ were at $z\sim 6$
is unclear, although there appear to be a plateau 
from $M_z=-18$ to $M_z=-22$ (Figure 9 of Popesso \etal 2004)
at the present time.
The uncertainty is of course that we do not know
when these large, old galaxies formed.
The largest possible upward correction factor would be
about $10$ by integrating the observed LF (Figure 9 of Popesso \etal 2004)
to the bright end, if all brighter galaxies were formed at the same time
as the dwarf galaxies.
Fourth, the ionizing photon escape fraction may be higher
than $10\%$ assumed, where the largest possible upward correction is $10$.
Fifth, if the IMF at $z\ge 6$ in these galaxies
is somewhat more top-heavy than the normal Salpeter IMF,
the estimated number of ionizing photons
would be increased correspondingly,
where the largest possible upward correction is about $30$
(Bromm, Kudritzki, \& Loeb 2001) for IMF dominated by 
metal-free massive stars.
Finally, quasars may contribute to the production of ionizing photons,
although their contributions depend crucially on the slope
of the quasar luminosity function (e.g, Fan \etal 2001; Wyithe \& Loeb 2002).
However, if star formation takes place mostly in proto-clusters,
quasar formation is likely to take place in these regions as well,
perhaps in denser parts of them.
In conclusion, it is physically plausible that we 
have underestimated the number of ionizing photons
from galaxies in proto-clusters by a factor of $\sim 5$ or more.

Perhaps more convincing is that a self-consistent reionization picture  
demands that the combination of these uncertain  
factors would increase the total number of ionizing photons
by a factor of $\sim 5$.
In this case, galaxies in proto-clusters become  
responsible for reionizing the entire universe at $z\sim 6$.
Only then one could naturally explain the lack of 
dwarf galaxies in the fields (Tully \etal 2002).
While some larger galaxies in regions of proto-clusters
may have also formed prior to reionization,
most larger galaxies in the fields should have
lagged in formation time compared to dwarfs in the fields
and therefore must form subsequent to reionization
in the hierarchical structure formation model.

In summary, we show that the number of ionizing photons 
derived from the directly observed dwarf galaxies in present galaxy clusters
should at least be able to ionize
$\sim 20\%$ of the IGM,
and we argue that the most self-consistent picture would be
that all ionizing sources in proto-clusters together
are able to reionize the entire IGM,
either because we have underestimated the overall star formation rates
in proto-clusters or because we have underestimated
the number of ionizing photons that enter the IGM (or both).

Is the production rate of ionizing photons from these
dwarf galaxies high enough to maintain ionization?
Using the estimated total mass of the 69 RASS-SDSS clusters
and assuming that they represent a typical 20\% of the mass
in their survey volume,
we find that the photon production rate from these faint galaxies is 

\noindent $5.6\times 10^{50} (f_{esc}/0.1) (t_{SB}/1\times 10^8{\rm yrs})$~s$^{-1}$~Mpc$^{-3}$,

\noindent where $t_{SB}$ is the assumed starburst duration,
which should be compared to the required rate 

\noindent $2.5\times 10^{51} C_{30} ({1+z\over 7})^3 
({\Omega_b h_{70}^2\over 0.04})^2$~s$^{-1}$~Mpc$^{-3}$

\noindent 
(Madau, Haardt, \& Rees 1999),
where $C_{30}=C/30$ and $C$ is the clumping factor of recombining gas.
Note that $C_{30}$ is likely to fall below
unity at $z\sim 6$ (Gnedin \& Ostriker 1997).
We see that the ionizing photon production rate from dwarf galaxies alone
falls short by a factor of $\sim 5$ for maintaining the global ionization.
Coincidentally, the fact that a factor of $5$ more ionizing photons than the adopted
value from the known dwarf galaxies in clusters
are needed to ionize every hydrogen atom in the IGM
suggests that once the entire IGM is reionized,
it can be kept ionized.

\section{Implications For Reionization Observations}

If galaxies (and quasars, possibly) in proto-clusters are primarily 
responsible for reionizing the universe at $z\sim 6$,
there are some very interesting implications.

First, it is likely that reionization starts from the centers
of these proto-clusters and proceeds in an inside-out fashion
(i.e., expanding in more or less spherical fashion).
Direct simulations (e.g., Sokasian \etal 2003; Ciardi \& Madau) show
fat-filaments-like H II regions.
However, simulated boxes are typically small, $\sim 10-20$Mpc,
smaller than or comparable to 
the expected correlation length of proto-clusters;
therefore, larger boxes are necessary to 
characterize the morphology of H II regions during reionization.
Within each individual H II region the ionization process
may progress in an outside-in way, as advocated by Miralda-Escud\'e,
Haehnelt \& Rees (2000) and Gnedin (2000).

Second, strong clustering of sources sitting on patches of individual
sizes of $10-30h^{-1}$Mpc would collectively 
produce very large individual H II regions of initial sizes
of $\sim 10-30$Mpc with some growing to $\sim 100$Mpc just before
overlapping with others.
This would have significant implications for future radio 
(e.g., Furlanetto, Zaldarriaga, \& Hernquist 2004b;
Kohler \etal 2005)
and CMB experiments (e.g., Knox, Scoccimarro, \& Dodelson 1998;
Santos \etal 2003; 
McQuinn \etal 2005; Zahn \etal 2005),
because these H II regions could grow to sizes
even larger than those expected from luminous quasars.
Under the assumption that ionizing photon production rate
per baryon does not depend on cluster richness,
we may translate the cluster mass function (Bahcall \& Cen 1993)
to an H II region size (cumulative) function (SF) at complete overlap:

\begin{equation}
n(>R) = 4\times 10^{-5} (R/R^*)^{-1} \exp(-R/R^*)~~h^3\hbox{Mpc}^{-3},
\end{equation}

\noindent
where $R^*=13.04h^{-1}$Mpc comoving and $R$ is the comoving radius
of an H II region. 
At earlier times the SF has the same
functional form with $R^*$ being smaller as
$R^*=13.04 f^{1/3}$Mpc,
where $f$ is the fraction of the IGM occupied by H II regions.
It is important to note that these H II regions 
centered on each proto-clusters are strongly clustered and
mergers between them will be an essential feature.
Therefore, a statistical treatment of the H II regions
is required.
The clustering of rich clusters of galaxies is mainly statistical,
determined by Gaussian statistics of peaks,
and a statistical treatment given by Kaiser (1984) 
can accurately describe their clustering properties, independent of redshift.
Empirical formulae such as that of Bahcall (1988), $r_0=0.4d$,
can also be used to described the clustering in simpler forms.
In combination with Equation (1) one will be able to 
compute various properties of H II regions during
reionization. Temporal evolution of $R^*$ (Equation 1)
would depend on astrophysics of galaxy formation but may be
computed empirically by matching the mean H II filling factor (i.e., $f$),
determined by independent observations such as from CMB observations.

The abundance of (richness 0 and above) clusters of galaxies is roughly
$10^{-5}$Mpc$^{-3}$ (Bahcall \& Cen 1993), 
which is about four orders of magnitude more abundant than
the luminous SDSS quasars (Fan \etal 2004).
Assuming a lifetime of quasars of $3\times 10^7$~yrs,
the abundance of SDSS quasars are still lower than
richness 0 and above clusters by a factor of $\sim 300$.
Thus, the luminous SDSS quasars
may be hosted by very rich proto-clusters with richness 3 or above
(with mean separation greater than $\sim 200h^{-1}$Mpc) and
perhaps today reside in the centers of these clusters,
if one believes that they form in the highest density peaks
in the hierarchical structure formation model.
We note that at complete overlap, the abundance of
SDSS quasars (taking into account the quasar lifetime)
is comparable to the abundance of H II regions of radius
$\sim 50h^{-1}$Mpc, which is comparable
to and somewhat larger than 
the sizes of SDSS quasar Stromgren spheres (e.g., Haiman \& Cen 2002),
having some important 
effects on the interpretation of SDSS quasar
observations (Mesinger \& Haiman 2004;
Wyithe \& Loeb 2004),

Third, the average distances between clusters of galaxies
range from $d=40h^{-1}$Mpc (for richness 0 above clusters) to 
$200h^{-1}$Mpc (for richness 3 and above clusters),
and the correlation length $r_0$ may be related to 
$d$ by $r_0=0.4d=16.0-80.0h^{-1}$Mpc (Bahcall 1988).
The mass correlation length at $z=6$ in the standard 
WMAP normalized CDM model (Spergel \etal 2003) is $r_M(6)=0.8$.
Therefore, these dwarf galaxies should be 
highly biased with a bias factor of $b\sim [r_0/r_M(6)]^{1.8/2}=15-63$,
consistent with recent observational indications (Malhotra \etal 2005;
Stiavelli \etal 2005).

Fourth, the large separations of ionizing sources 
suggest that there should be large spatial fluctuations of 
neutral fraction in the IGM up to 
cluster scale of $\sim 100$Mpc.
It is intriguing to notice that,
if one were to interpret
the difference between the highly ionized regions suggested
by SDSS J1148+5251 and significantly neutral regions around
SDSS J1030+0524 as spatial variations
(Oh \& Furlanetto 2005),
our model may provide a natural explanation.

Finally, the large H II regions would enable the flux
of a large fraction of the embedded sources to be transmitted without
significant attenuation.
With present observational sensitivities
some of the more luminous sources among them 
will then become individually visible
even prior to the completion of overlap of individual H II regions
This may provide a more natural explanation for the
observed high redshift galaxies prior to complete reionization 
(Hu \etal 2002; 
Cuby \etal 2003;
Rhoads \etal 2004;
Stern \etal 2005; Taniguchi \etal 2005)
(see Haiman 2002 for an alternative explanation);
this point was made by Furlanetto, Hernquist, \& Zaldarriaga (2004)
and Wyithe \& Loeb (2005) in a more general
context of source clustering in cold dark matter model.
A definitive check is this:
when more sensitive observations become available,
one should expect to see clusters of fainter sources
in the vicinity of known galaxies at $z>6$.
As a corollary, the lack of evolution of \lya emitting galaxies
from $z=6.5$ to $z=5.7$ will not be unexpected (Malhotra \& Rhoads 2004).
In fact, it might be expected that the observed (and true)
LF of \lya emitting galaxies
may fall from $z=6.5$ to $z=5.7$ due to suppression
of star formation, especially at the low luminosity end after reionization.
While some of their aged, likely stellar only counterparts are seen today
as dwarf galaxies in X-ray clusters of galaxies,
clearly, these dwarfs galaxies in starburst phase at $z\ge 6$ when 
they produce a significant number of ionizing photons  
are just beyond reach 
with the current deepest surveys (Bouwens \etal 2004; Bunker \etal 2004).
But we emphasize this:
we need to go deeper and go higher redshift beyond reionization
to find these sources; the abundance of lower redshift 
counterparts subsequent to reionization may be masked by
physics of reionization.

\section{Conclusions and Discussion}

We show that faint dwarf galaxies, now seen in clusters of galaxies,
when they form at $z\ge 6$,
are able to produce ionizing photons at a sufficient rate
to be responsible for reionizing at least $20\%$ of the 
universe at $z\sim 6$.
We argue that several uncertain factors may allow for the possibility
that galaxies in proto-clusters reionize the entire universe at
$z\sim 6$,
only under which one may
construct a self-consistent reionization picture.
The implications are substantially profound
and observationally verifiable (see \S 3).
In particular, fluctuations of neutral fraction
across very large spatial dimensions ($\sim 100$Mpc) 
are expected, so are very large H II regions  of size $\sim 100$Mpc.
We give an H II region size function in \S 3 (Equation 1).
Such large H II regions may be observable 
by upcoming radio (such as LOFAR, MWA, PAST and SKA)
and CMB experiments (WMAP and PLANCK).

Furlanetto and collaborators have investigated
extensively the growth of large H II regions 
based on the Extended Press-Schechter theory,
having the advantage of being able to treat
scales larger than possible in present simulations
(Furlanetto, Zaldarriaga, \& Hernquist 2004a;
(Furlanetto, Hernquist, \& Zaldarriaga 2004;
Furlanetto \& Oh 2005).
What is interesting is that these two completely
different and complementary approaches,
ours based on direct observations and theirs on
the cold dark matter model,
appear to converge onto a similar conclusion
that reionization process likely proceeds inside-out 
on large scales and is very inhomogeneous on scales up to 
many megaparsecs.

Within such a physical picture of cosmological reionization,
a proper numerical computation would require a box size 
of at least $100$Mpc, in agreement with theoretical 
considerations based on clustering properties 
in the cold dark matter model (Barkana \& Loeb 2004).
This is a formidable challenge to workers in the field,
especially in the context of radiation hydrodynamic simulations.

Observations evidently paint a picture where 
dwarfs form much more abundantly in proto-clusters
than in the fields, for which Tully \etal (2002)
give a convincing theoretical demonstration 
in the context of cold dark matter model.
Here we will give a simple argument that,
in order to form a self-consistent picture,
the duration of the final reionization 
has a redshift interval of $\Delta z=1-2$,
as follows.
Proto-cluster regions at redshift $z=6$
have a linear overdensity of $\delta_c/D(0)=0.18\delta_c$,
where $D(0)$ is the linear growth factor from $z=6$ to $z=0$
in the standard CDM model (Spergel \etal 2003) and
$\delta_c$ is the linear critical overdensity for collapse.
Therefore, roughly speaking, 
halo formation in the fields (where overdensity is close to zero)
lags that in proto-cluster regions 
by a redshift interval $\Delta z = 0.18\delta_c/\delta_c \times (1+6) = 1.3$.
The fact that formation of dwarf galaxies in the fields 
was largely suppressed suggests that
reionization did not take significantly longer than that interval $\Delta z$.

\acknowledgments
I thank Andrey Kravtsov for useful discussion
and Steven Furlanetto for helpful comments.
This research is supported in part by 
NAG5-13381, AST-0206299 and AST-0407176.

\end{document}